\documentclass[12pt]{iopart}
\usepackage{iopams}
\usepackage{graphicx}
\usepackage{hyphenat}
\usepackage{multirow}
\usepackage{color}
\usepackage[utf8]{inputenc}

\renewcommand{\theta}{\vartheta}

\begin{document}

\title{Gravity-induced dynamics of a squirmer microswimmer in wall proximity}

\author{Felix R\"uhle, Johannes Blaschke, Jan-Timm Kuhr and Holger Stark}

\address{Technische Universit\"at Berlin, Institut f\"ur Theoretische Physik, Hardenbergstr. 36
D-10623 Berlin, Germany}
\ead{ruehle@tu-berlin.de}
\vspace{10pt}

\begin{abstract}
We perform hydrodynamic simulations using the method of multi-particle collision dynamics and a theoretical analysis to study a single squirmer microswimmer at high P\'eclet number, which moves in a low Reynolds number fluid and under gravity.
The relevant parameters are the ratio $\alpha$ of swimming to bulk sedimentation velocity and the squirmer type $\beta$. The combination of self-propulsion, gravitational force, hydrodynamic interactions with the wall, and thermal noise leads to a surprisingly diverse behavior. At $\alpha > 1$ we observe cruising states, while for $\alpha<1$ the squirmer resides close to the bottom wall with the motional state determined by stable fixed points in height and orientation. They strongly depend on the squirmer type $\beta$. While neutral squirmers permanently float above the wall with upright orientation, pullers float for $\alpha$ larger than a threshold value $\alpha_{\mathrm{th}}$ and are pinned to the wall
below $\alpha_{\mathrm{th}}$. In contrast, pushers slide along the wall at lower heights, from which thermal orientational fluctuations drive them into a recurrent floating state with upright orientation, where they remain on the timescale of 
orientational persistence.
\end{abstract}

\pacs{47.63.Gd,47.63.mf,47.57.ef}
%
\vspace{2pc}
\noindent{\it Keywords}: low-Reynolds-number flows, microswimmer dynamics, swimming under gravity, hydrodynamic wall interactions
%
%
%
%

\section{Introduction}
\label{intro}

The fact that active particles are inherently in non\hyp{}equilibrium has stimulated experimental \cite{HowseGolestanian2007,JiangSano2010,ZhangSwinney2010,HerminghausBahr2014}, theoretical \cite{Ramaswamy2010,MarchettiSimha2013} and numerical  
\cite{SaintillanShelley2007,SaintillanShelley2008,IshikawaLocseiPedley2008,ZoettlStark2016} research in the last decade. 
This is also true for fluid systems at low Reynolds number, where swimmers on the micron scale are considered, i.e.\ bio\-logical organisms \cite{LaugaPowers2009,BaskaranMarchetti2009} and synthetic particles \cite{ButtinoniSpeck2013,SimmchenSanchez2016} as well as continuum models 
thereof \cite{SimhaRamaswamy2002}. A decisive factor for such 
microswimmers are hydrodynamic interactions with surfaces and with each other~\cite{BerkeLauga2008,LLopisPagonabarraga2006,LlopisPagonabarraga2010,ZoettlStark2014,SchaarStark2015,BlaschkeStark2016,ElgetiGompper2015,Guzman-LastraLoewen2016,LintuvuoriMarenduzzo2016,LiArdekani2014,UspalTasinkevych2015,UspalTasinkevych2015rheotaxis}.

The behavior of active particle systems is intriguing and often counter-intuitive. This is especially true when considering collective dynamics. 
For example, one can find motility-induced phase separation with purely repulsive particle\hyp{}particle interactions at low densities where passive particles would not phase-separate~\cite{ButtinoniSpeck2013,BialkeSpeck2013,TailleurCates2008,FilyMarchetti2012,RednerBaskaran2013,CatesTailleur2015}.
In bacterial systems the formation of biofilms~\cite{OTooleKolter2000} has deservedly attracted much attention. 
However, even the trajectory of a single active agent in a solvent can be very interesting. \emph{E.coli} bacteria swim in circles 
close to boundaries~\cite{LaugaStone2006} and the sperm cell's navigation under flow has intrigued
researchers for over 50 years~\cite{BrethertonRothschild1961,JikeliKaupp2015}.
Recently, the swinging and tumbling trajectories of single active particles in Poiseuille flow have been classified~\cite{ZoettlStark2012,ZoettlStark2013} 
and swinging has been observed for the African typanosome, a parasite causing the sleeping sickness~\cite{UppaluriPfohl2012}.

An ongoing field of study is the question how the addition of external forces
influences the force-free propulsion of active particles. An example are self-propelled particles or particle chains with 
additional magnetic moments that alter the effective diffusion constant~\cite{FanSandoval2017} or give rise to new interesting features such as
bifurcations and instabilities of linear molecules or rings made of microswimmers~\cite{Guzman-LastraLoewen2016,BabelMenzel2016}.
A very natural influence to consider is gravity.  Breaking translational symmetry along one spatial direction leads to bound swimmer states, polar order, and fluid pumps~\cite{DrescherGoldstein2009,PalacciBocquet2010,EnculescuStark2011,HennesStark2014}. 
Furthermore, appealing pattern formation of bacteria occur, known as bioconvection~\cite{PedleyKessler1992}. Novel phenomena have been discovered such as gravitaxis of asymmetric swimmers~\cite{tenHagenBechinger2014}, inverted sedimentation profiles of bottom-heavy swimmers~\cite{WolffStark2013}, the formation of thin phytoplankton layers in the coastal ocean~\cite{DurhamStocker2009}, and 
rafts of active emulsion droplets, which potentially occur due to phoretic interactions~\cite{KruegerMaass2016}.

Spherical squirmers mimic ciliated organisms like the Volvox algae or are used as model swimmers to explore the 
consequences of their self-generated flow fields \cite{Lighthill1952,Blake1971}. 

Recently, states of squirmers close to a bounding wall have been presented \cite{LintuvuoriMarenduzzo2016,LiArdekani2014}. In Ref.\ \cite{LintuvuoriMarenduzzo2016} also a short-range repulsion from the wall was included, which lead to oscillatory variations of the height above the
wall with a mean distance close to one particle radius. It was also demonstrated that far-field hydrodynamics cannot fully explain the observed phenomenology.

In this article we report on full hydrodynamic simulations of a single squirmer under gravity close to bounding walls and supplement it by a theoretical analysis. In particular, we concentrate on the case where the squirmer speed is comparable to the bulk sedimentation velocity. We find that this setting suffices to create very diverse and unforeseen novel dynamics on distances several squirmer radii away from a bottom wall. To guide the reader, we first introduce the main phenomenology observed in our simulations.

\subsection{Phenomenology}
\label{subsec.phenom}

\begin{figure}
\centering
 \resizebox{0.45\textwidth}{!}{
\includegraphics{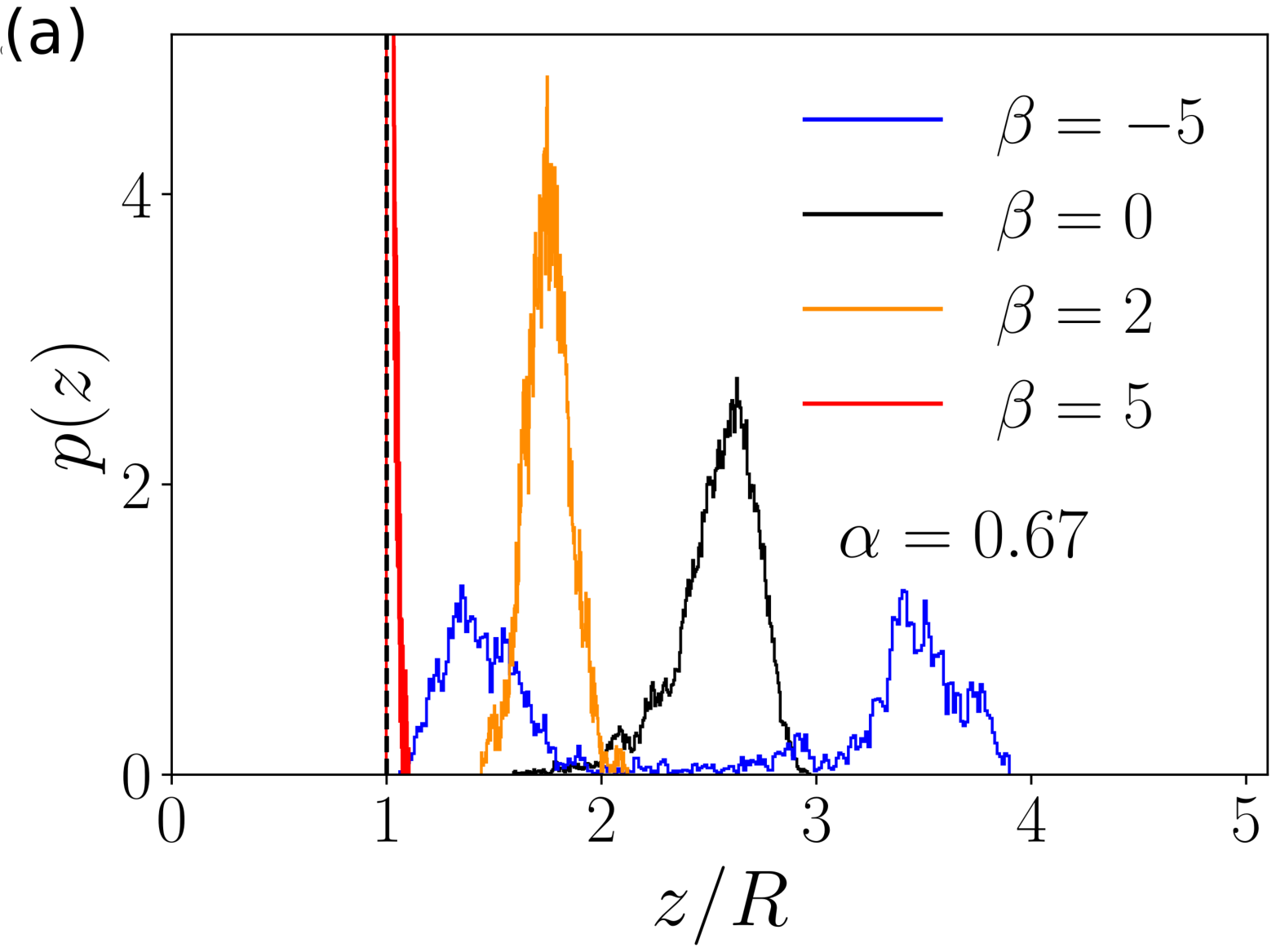}}
\resizebox{0.45\textwidth}{!}{
\includegraphics{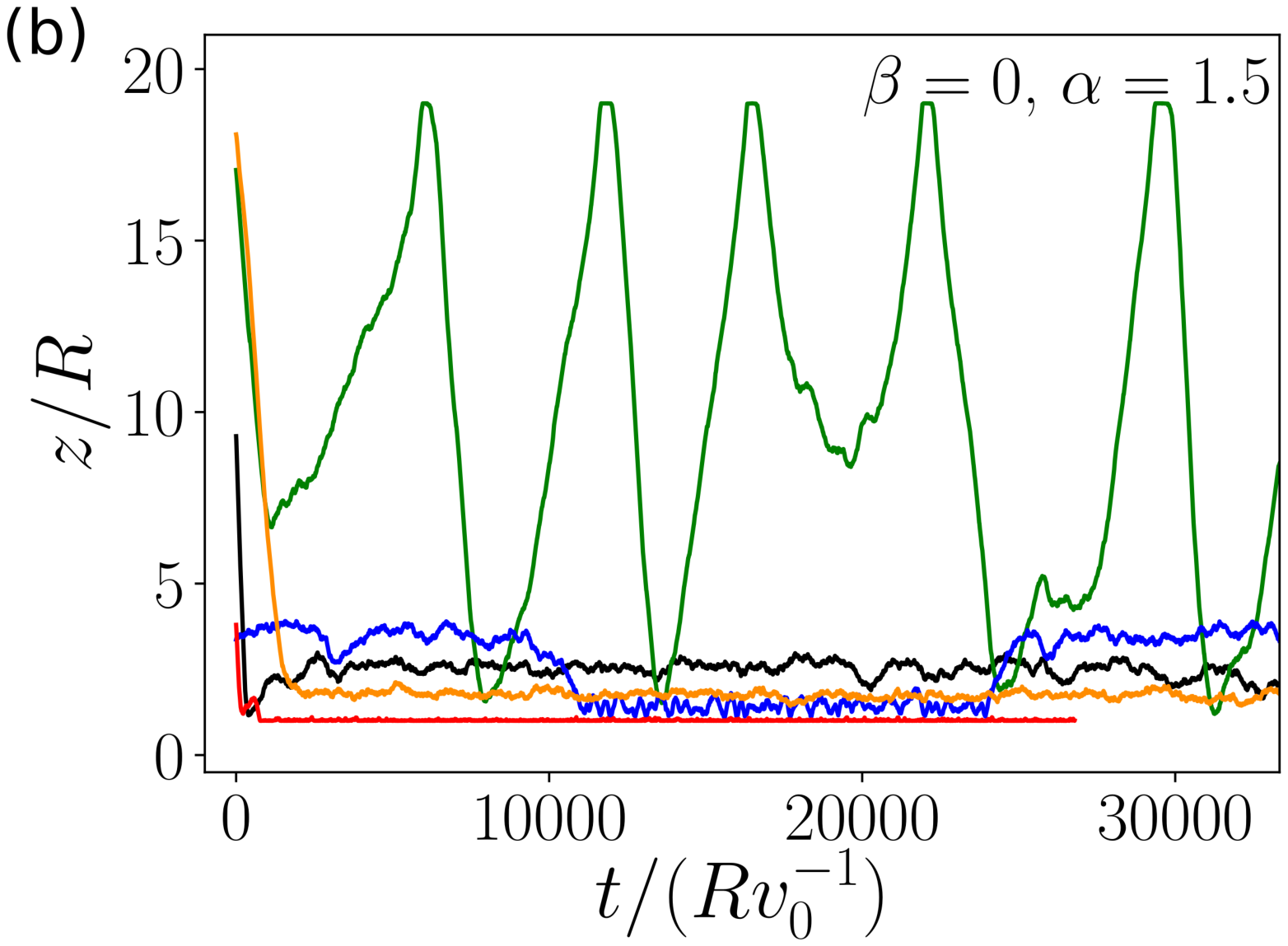}}

\caption{
a) Distribution $p(z)$ of heights $z$ during motion of a squirmer for $\alpha = v_0/v_g=0.67$ for different squirmer types $\beta$ illustrating stable floating ($\beta=0,2$), wall pinning ($\beta=5$), and the bimodal state of
recurrent floating and sliding ($\beta=-5$). b)
Corresponding trajectories $z(t)$. 
The green line illustrates a cruising state for $\alpha=1.5$ and $\beta=0$.}
\label{fig:hist}
\end{figure}

We will use the method of multi-particle collision dynamics to
simulate a single spherical squirmer moving under gravity in a quiescent fluid bounded by a top and bottom wall.
We will demonstrate that already such a simple setting shows different motional states. We shortly summarize them here.
The squirmer propels itself with a velocity $v_0$ due to a tangential surface velocity field, which is controlled by the squirmer-type parameter $\beta$ and thereby allows us to distinguish between pullers ($\beta>0$), neutral squirmers ($\beta=0$), and pushers ($\beta < 0$). 

In the following, the ratio 
\begin{equation}
\alpha = v_0/v_g
\end{equation}
of the swimming velocity $v_0$ and the bulk sedimentation velocity $v_g$ will be the relevant parameter, while all squirmers move persistently with a large P\'eclet number.

A neutral squirmer with $\alpha >1$, where self-propulsion dominates, continuously cruises between the top and bottom wall [see Fig.\ \ref{fig:hist}(b) and video M1 in the supplemental material]. Each time it reaches a wall, its orientation is reversed so that it moves persistently to the other wall.

However, we will mainly concentrate on the case $\alpha < 1$, in particular, where gravity and activity are comparable to each other.
Then the squirmer resides close to the bottom wall but its motional state dramatically depends on the squirmer type $\beta$, as we illustrate for $\alpha=0.67$ in Fig.\ \ref{fig:hist}.
The neutral squirmer and a weak puller ($\beta=2$) show stable floating in a finite distance above the bottom wall, where the maximal reachable height is 
larger for the neutral squirmer (see also video M2 in the supplemental material). 
The strong puller ($\beta=5 $), however, is in a wall-pinned state and hardly escapes the wall at all.  
Finally, the behavior of a strong pusher ($\beta=-5$) is strikingly different.
It recurrently switches between floating at heights larger than a neutral squirmer
and sliding along the wall at lower heights (see also video M3 in the supplemental material). 
Particularly the fact that long-lived states can occur several radii away from the wall (see Fig.\ \ref{fig:hist}) strikes us as a most interesting feature. While hovering states have been reported in connection with catalytically active particles \cite{UspalTasinkevych2015}, squirmers near walls either show, for example, the already mentioned oscillatory near-wall dynamics or escape the wall altogether \cite{LintuvuoriMarenduzzo2016,LiArdekani2014}.
In this article we will analyze and explain in detail all the motional states illustrated in Fig.\ \ref{fig:hist} and videos M1-M3 in order to obtain a full understanding of the motional states of a squirmer under gravity and close to a bounding bottom wall. This will serve as a reference case for future studies.

The article is organized as follows. In Sec.\ \ref{sec.methods} we outline the squirmer model and the simulation technique of multi-particle collision dynamics. In Sec.\ \ref{sec.theory} the theory of a squirmer under gravity and its hydrodynamic interactions with a wall are discussed
and first conclusions for the observed squirmer orientations are drawn.
We continue with Sec.\ \ref{sec.discussion}, where we first present our simulation results and then discuss them further in the light of theory.
We conclude in Sec.\ \ref{sec.conclusion}.

\section{Squirmer Model and Simulation Method}
\label{sec.methods}

The spherical squirmer is a versatile model system for active swimmers, such as various bacteria and artificial microswimmers like Janus particles \cite{Lighthill1952}. 
Its motion is induced by a tangential flow field on the squirmer surface 
\cite{Blake1971,IshikawaPedley2006},
\begin{equation}
\label{eq:surface_field}
\mathbf{v}_s(\mathbf{r}_s) = B_1 \left( 1 + \beta \hat{\mathbf{e}} \cdot \hat{\mathbf{r}}_s \right) \left[ \left( \hat{\mathbf{e}} \cdot
\hat{\mathbf{r}}_s \right) \hat{\mathbf{r}}_s - \hat{\mathbf{e}} \right].
\end{equation}
Here,
$\hat{\mathbf{e}}$ is the squirmer orientation and 
$\hat{\mathbf{r}}_s = \mathbf{r}_s / | \mathbf{r}_s | $, where $\mathbf{r}_s$ is a spatial vector pointing from the center to the squir\-mer surface. 
In the following, we only take into account the first two modes of the Fourier
expansion used in Refs. \cite{Lighthill1952,Blake1971} for the surface flow field, $\mathbf{v}_s(\mathbf{r}_s)$, $B_1$ and $B_2$.
The squirmer's swimming velocity is determined by $v_0 = 2/3 B_1$ and the parameter $\beta = B_2 / B_1$ characterizes the squirmer type, as introduced above.
Far from the squirmer surface the two first modes create the velocity fields of a source dipole~($\sim r^{-3}$)  and force dipole~($\sim r^{-2}$), respectively \cite{SpagnolieLauga2012}. 

Our simulations should account for the full hydrodynamics at low Reynolds numbers. 
Thus, we use the meso\-scale particle-based method of multi-particle collision dynamics (MPCD) \cite{MalevanetsKapral1999,NoguchiGompper2007}
to solve the Stokes equations. The details of our implementation of MPCD follows our previous works in
Refs.\ \cite{ZoettlStark2014,BlaschkeStark2016}. We only give a few details here.
The fluid is modeled by approximately $5 \cdot 10^5$ point particles of mass $m_0$, 
the positions and velocities of which are updated in two consecutive steps. In the streaming step each fluid particle moves with its velocity during time $\Delta t$. Thus, fluid momentum flows in the simulation box but 
is also transferred to the squirmer, when the fluid particles collide with it.
In the collision step fluid particles are sorted into cubic cells of side length $a_0$. Then, the velocities of the fluid particles are modified by a collision operator, for which we use the MPC-AT+a rule \cite{NoguchiGompper2007}. 
It conserves total momentum and angular momentum of the fluid particles in each cell and sets up a thermostat at temperature $T_0$. Importantly, momentum conservation is necessary for recovering the Navier-Stokes equations on the length scale of the mean free path of a fluid particle \cite{NoguchiGompper2007,GoetzeGompper2007,NoguchiGompper2008}. Additionally, the method includes thermal noise. 
Note that the collision cells need to be shifted for each new collision step to restore Galilean invariance \cite{IhleKroll2003}. At surfaces the so-called bounce-back rule is applied to the fluid particles, which implements the no-slip boundary condition at bounding walls and the flow field of eq.\ (\ref{eq:surface_field}) at the squirmer surface \cite{ZoettlPHD}. Squirmer dynamics is resolved during the streaming step by 20 molecular dynamics steps, where we also include the gravitational force.

Hydrodynamic flow fields and near- and far-field interactions of squirmers  are well reproduced by the MPCD method \cite{ZoettlStark2014,GoetzeGompper2010,DowntonStark2009}. We set the squirmer radius to $R=4a_0$ and the leading surface velocity mode to $B_1 = 0.1$ (in MPCD velocity units $\sqrt{k_BT_0/m_0}$).
Since we choose for the duration of the streaming step $\Delta t = 0.02 a_0\sqrt{m_0/k_BT_0}$, we have for the fluid viscosity 
$\eta = 16.05 \sqrt{m_0k_BT_0}/a_0^2$ \cite{NoguchiGompper2008,ZoettlPHD}. 
The translational and rotational thermal diffusivities in bulk fluid then become 
$D_T = k_BT /(6\pi\eta R) \approx 8\cdot 10^{-4} a_0\sqrt{k_BT_0/m_0}$  and 
$D_R = k_BT /(8\pi\eta R^3) \approx 4*10^{-5}\sqrt{k_BT_0/m_0}/a_0^2$, respectively. 
With $v_0 = 2/3 B_1$ this yields the active P\'{e}clet number $\mathrm{Pe} = Rv_0 / D_T = 330$ and the persistence number $\mathrm{Pe}_r= v_0/(RD_R)=420$.
The simulation box has an edge length of $20 R$ in $x$-, $y$- and $z$-direction. 
While it is bounded by a top and bottom no-slip wall, we use periodic boundary conditions in the horizontal plane.

\section{Theory}

\label{sec.theory}

\begin{figure} 
\centering
\resizebox{0.27\textwidth}{!}{
 \includegraphics{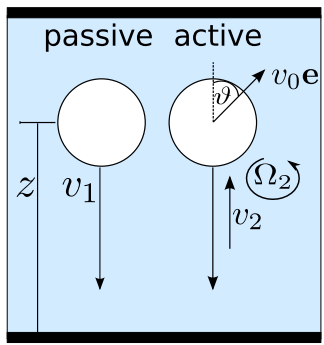}
}
\caption{A passive particle sediments with a velocity $v_1$ due to the height-dependent friction coefficient. An active particle, in addition, moves with the swimming velocity $v_0 \cos \theta$ along the vertical. Its self-generated flow field interacts with the wall and induces a deterministic linear ($v_2$) and angular ($\Omega_2$) velocity.}
\label{vcontrib}      
\end{figure}

In wall proximity a squirmer experiences three  deterministic contributions to its vertical velocity (see Fig.\ \ref{vcontrib}). First, it self-propels with velocity $v_0 \cos \theta$, where $v_0$ is the swimming velocity along the orientation vector $\mathbf{e}$ and $\theta$ the angle against the normal. Second, it 
sediments with a height-dependent velocity $v_1$ since the friction coefficient depends on the height $z$ above the wall, which represents the hydrodynamic interaction of a passive particle with the wall. Third, its self-generated flow field 
also hydrodynamically interacts with the wall and thereby induces a linear ($v_2$) and angular ($\Omega_2$) velocity.

We therefore write for the total vertical velocity and total angular velocity,
\begin{equation}
\label{eq:3velocities}
v = v_0 \cos\vartheta - v_1 + v_2 \quad \mathrm{and} \quad \Omega = \Omega_2 \, .
\end{equation}
For a passive particle all but the term $v_1$ would vanish. 
Stochastic motion due to translational and rotational diffusion are not considered here. 
For the deterministic system 
\begin{equation}
\label{eq:dynamical_system}
\left(\begin{array}{c}
\dot{z}\\
\dot{\theta}\\
\end{array}\right) = \left(\begin{array}{c}
v\\
\Omega\\
\end{array}\right) =: f(z,\vartheta) \,
\end{equation}
where the equilibrium states at $(z^*,\theta^*)$ (fixed points) follow from $f(z^*,\theta^*) = 0$, one can then identify the stable states by performing a stability analysis and demanding the eigenvalues of the Jacobian $Df(z^*,\vartheta^*)$  to be negative. This procedure should, in principle, identify the motional states introduced in Fig.\ \ref{fig:hist}.
However, we can only perform this stability analysis in the far-field approximation explicitly. This does not identify all the observed motional states as we will discuss below.

In the following, we explain the different contributions $v_1$, $v_2$, and $\Omega_2$ in more detail in Secs.\ \ref{height_friction}, \ref{hi_squirmer_walls}. Then, we first comment on stable squirmer orientations in wall proximity using far-field and lubrication expressions in Secs.\ \ref{subsec.stable_orient}.
Thereby, we will obtain a first understanding of the motional states presented in Fig.\ \ref{fig:hist}.
We complete the far-field analysis of the dynamical system in Sec.\ \ref{floating_heights}, where we address
stable squirmer heights.

\subsection{Height-dependent sedimentation velocity}
\label{height_friction}

The squirmer experiences a gravitational force $\mathbf{F} = -mg \mathbf{\hat{z}}$, where in a real experiment $g = g_0 (1 - \rho_\mathrm{f}/\rho_\mathrm{p})$ 
depends on the mismatch of fluid and particle densities $\rho_\mathrm{f,p}$ and $g_0$ is the gravitational acceleration. 
Tuning $g$ by tuning the solvent density $\rho_\mathrm{f}$, has been applied in Ref.\ \cite{KruegerMaass2016} 
to active emulsion droplets.

The height-dependent sedimentation velocity 
\begin{equation}
v_1(z)  =  \frac{mg}{ \gamma(z) }
\end{equation} 
is now determined by a height-dependent friction coefficient, which takes the bounding walls into account.
For one wall its inverse can be written as an expansion up to third order in $R/z$ \cite{Wakiya1960,PerkinsJones1992}:
\begin{equation}
\label{eq:vsed_1wall}
\gamma_\mathrm{\mathrm{1w}}^{-1}(z) \approx 
\gamma_\infty^{-1} \left[1 - \frac{9}{8}\frac{R}{z} + \frac{1}{2}\left(\frac{R}{z}\right)^3 \right] \, .
\end{equation}
Here $\gamma_\infty = 6\pi \eta R$ is the Stokes friction coefficient of a particle with radius $R$ in a bulk fluid with shear 
viscosity $\eta$. Note that close to a wall friction becomes anisotropic and in eq.\ (\ref{eq:vsed_1wall}) only the 
component perpendicular to the wall is considered.

The friction coefficient in eq.\ (\ref{eq:vsed_1wall}) is only valid for a single wall. To model our simulation results, we use a
simple approximation for the two-wall coefficient:
\begin{equation}
\label{eq:twowalllambda}
\gamma_\mathrm{2w}^{-1} = \gamma_\mathrm{1w}^{-1}(z) + \gamma_\mathrm{1w}^{-1}(h-z) -1
\end{equation}
where the first and second term on the left-hand side refer to the single-wall friction coefficients oft the bottom and top wall, respectively, and $h$ is the box height. Obviously there is an error connected with this procedure,
calculated in Refs. \cite{BhattacharyaBlawzdziewicz2002,Jones2004,BhattacharyaWajnryb2005} to be 15\%-18\%.

The height-dependent friction coefficient is already sufficient to understand the stable floating of a squirmer
close to the bottom wall in the lower half of the simulation box.
Suppose the upward swimming squirmer floats at a certain height where swimming velocity $v_0$ and sedimentation velocity cancel. If (thermal) fluctuations drive it to larger heights, the friction coefficient
decreases. As a result the sedimentation velocity increases and drives the squirmer back to the initial height. Similarly, fluctuations to smaller heights decrease the sedimentation velocity and the squirmer moves upwards. However, the flow field generated by the squirmer during its swimming motion also hydrodynamically
interacts with the bottom wall, so that the behavior depends on squirmer type $\beta$.

\subsection{Hydrodynamic interactions of squirmer flow field with a wall}
\label{hi_squirmer_walls}

In the following we only consider hydrodynamic squirmer-wall interactions due to the self-propulsion flow field of the squirmer. The effect of the gravitational force was treated in the previous section.

\subsubsection{Far field}

The velocity far field of the squirmer consist of a force dipole with strength $p$ and a source dipole with strength $s>0$:
\begin{equation}
\label{forcedipole}
\mathbf{v}(\mathbf{r}) = - \frac{p}{r^2} [1-3\left(\mathbf{e}\cdot\hat{\mathbf{r}}\right)^2 ] \hat{\mathbf{r}}
- \frac{s}{r^3} [\mathbf{e} - 3\left(\mathbf{e}\cdot\hat{\mathbf{r}}\right) \hat{\mathbf{r}} ] \, ,
\end{equation}
where $r=\vert\mathbf{r}-\mathbf{r}_0\vert$, $\mathbf{\hat{r}} = (\mathbf{r}-\mathbf{r}_0) / r$, and $\mathbf{r}_0$ is the position and 
$\mathbf{e}$ the orientation of the squirmer. The strengths $p$ and $s$ are connected to the squirmer velocity $v_0$ and type $\beta$~\cite{SpagnolieLauga2012,ZoettlPHD}:
\begin{equation}
\label{dpstrength}
p = -\frac{3}{4} \beta v_0 R^2 \quad \mathrm{and} \quad s = \frac{1}{2} v_0 R^3 \, .
\end{equation}
Note while $s>0$, the force dipole varies in the range $p\in(-\infty,\infty)$.

\begin{figure}
\centering
\resizebox{0.5\textwidth}{!}{
 \includegraphics{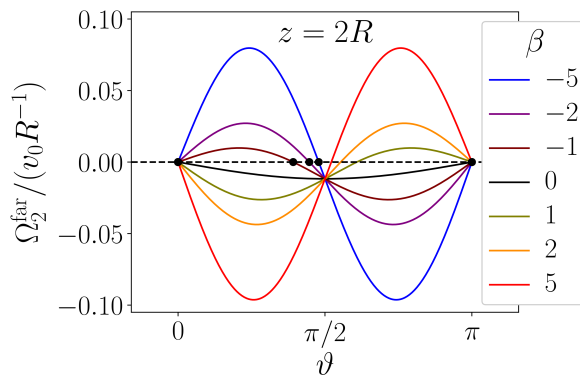}}
\caption{
Wall-induced angular velocity of a squirmer in far-field approximations, $\Omega_2$ of eq.\ (\ref{angular}), versus 
$\vartheta$ for different $\beta$ at $z=2R$.
Stable fixed points ($\Omega_2 =0$ and negative slope) are indicated by black dots.
}
\label{omega_ana_ff}
\end{figure}

The wall-reflexion fields for both dipoles in the far-field approximation are known. Therefore, the wall-induced linear ($v_2$) and angular ($\Omega_2$) velocities of the squirmer can be calculated from
Fax{\'e}n's theorem\ \cite{BerkeLauga2008,SpagnolieLauga2012}. Using Eqs.\ (\ref{dpstrength}), they are written as

\begin{eqnarray}
\label{eq:velocity_ff}
v_2  & = & \frac{v_0}{2}\left(\frac{R}{z}\right)^2 \left[\frac{9}{16}\beta\left(1-3\cos^2\theta \right)-\frac{R}{z}\cos\theta\right] \\
\label{angular}
\Omega_2 
& = & -\frac{v_0}{R} \frac{3}{16}  \left( \frac{R}{z} \right)^3
\sin\theta \left[\frac{3}{2}\beta \cos\theta + \frac{R}{z}\right]
\end{eqnarray}
Note that we defined $\Omega_2$ such that $d \theta / d t = \Omega_2$.
Figure\ \ref{omega_ana_ff} plots $\Omega_2$ versus orientation angle $\vartheta$ for different $\beta$ at $z=2R$. For increasing $z$ the stable fixed points in the middle ($\Omega_2=0$) move closer to $\pi/2$ and the overall strength of $\Omega_2$ decreases.

\subsubsection{Near field in lubrication approximation}

\begin{figure}
\centering
\resizebox{0.5\textwidth}{!}{\includegraphics{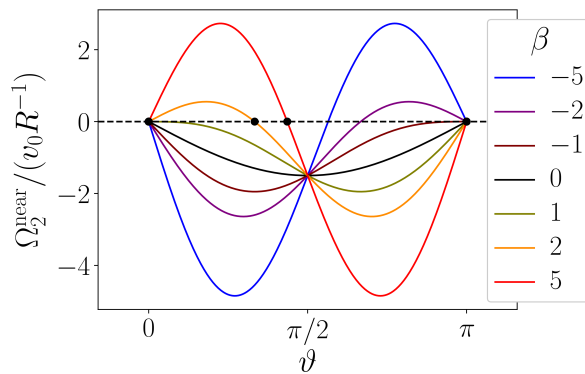}}
\caption{
Wall-induced angular velocity of a squirmer from lubrication theory, $\Omega_2$ of eq.\ (\ref{eq:angular_nf}), versus $\vartheta$ for different $\beta$.
Stable fixed points ($\Omega_2 =0$ and negative slope) are indicated by black dots.
}
\label{omega_ana_nf}
\end{figure}

In our simulations, squirmers also encounter the top or bottom wall, where far-field hydrodynamics does not apply. Therefore, we need to take into account results from lubrication theory, which gives for the wall-induced angular velocity~\cite{SchaarStark2015,LintuvuoriMarenduzzo2016,IshikawaPedley2006},
\begin{equation}
\label{eq:angular_nf}
\Omega_2 = \frac{3}{2}\frac{v_0}{R}
\sin\theta\left(\beta\cos\theta -1\right) +\mathcal{O} (1/\log(\varepsilon)),
\end{equation}
where $\varepsilon=(z-R)/R$ is the smallness parameter giving the reduced distance of the squirmer surface from a wall.
Figure\ \ref{omega_ana_nf} plots  $\Omega_2$ versus $\vartheta$ for different
squirmer types $\beta$.

Successful analytical methods tackling lubrication forces of self-propelled particles have been described quite recently \cite{Yariv2016,PapavassiliouAlexander2017}. Here, we do not attempt to calculate the vertical velocity in the near field. The authors of Ref. \cite{LintuvuoriMarenduzzo2016} showed that its leading order depends on longer-range interactions between squirmer and wall and are hence outside the scope of the lubrication approximation.

\subsection{Stable squirmer orientations}  
\label{subsec.stable_orient}

\renewcommand{\arraystretch}{2}
\begin{table}
\centering
\caption{Stable orientation angle $\vartheta^{*}$ for different squirmer types from
lubrication theory and in far-field approximation.
}
\label{tab:orientations}
\begin{tabular}{|c|c|c|}
\hline 
$\vartheta^{*} $ & lubrication & far field \\     
\hline 
\hline 
\multirow{2}{*}{pusher} & 0 & $\mathrm{acos} \left[ \frac{2}{3 |\beta |} \frac{R}{z} \right]$ if $\vert\beta\vert>\frac{2R}{3z}$ \\
 & $\pi$ if $\beta < -1$ &    0   otherwise $\!\!\!\!$ \\
\hline 
neutral & $0$ & $0$\tabularnewline
\hline 
\multirow{2}{*}{puller} & 0 if $\beta < 1$ & 0 \\
& $\mathrm{acos} \beta^{-1}$ if $\beta > 1$ & $\pi$ if $\beta>\frac{2R}{3z}$ \\[.5ex]
\hline 
\end{tabular}
\end{table}

We now calculate the stable squirmer orientations in far-field approximation and in the lubrication regime at the wall
by setting $\Omega_2(\theta^*) = 0$ in eqs.\ (\ref{angular}) and (\ref{eq:angular_nf}). In addition, the stability condition
$ \partial \Omega_2 / \partial \theta \vert_{\theta^*}  < 0$ has to be fullfilled. From eqs.\ (\ref{angular}) and (\ref{eq:angular_nf})
we obtain the respective derivatives as:
\begin{eqnarray}
\frac{\partial \Omega^{\mathrm{far}}}{\partial \theta} & \propto & -\frac{3}{2}\beta \left(\cos^2\theta - \sin^2\theta\right) + \frac{R}{z}\cos\theta  < 0 \\
\frac{\partial \Omega^{\mathrm{near}}}{\partial \theta} & \propto & - \cos\theta + \beta (\cos^2\theta - \sin^2\theta ) 
< 0 \, .
\end{eqnarray}
The stable orientation angles are indicated in Figs.\ \ref{omega_ana_ff} and \ref{omega_ana_nf} and summarized in Tab.\ \ref{tab:orientations}.

We shortly discuss the stable orientations and give a first understanding of the squirmer states close to a wall
as illustrated in Fig.\ \ref{fig:hist}. A thorough understanding of the squirmer dynamics is provided in 
Sec.\ \ref{sec.discussion}.
The neutral squirmer always points away from the wall ($\vartheta^{*}=0$) both when it is very close to the wall and in the far-field regime. This explains the cruising motion for large swimming velocity, $\alpha > 1$, introduced in Fig.\ \ref{fig:hist}b). Whenever the squirmer comes close to the wall, it reorients quickly due to hydrodynamic interactions with the wall and leaves. A stable upward orientation
near the bottom wall is also a necessary condition for the permanently floating squirmer introduced in Fig.\ \ref{fig:hist} for $\alpha < 1$. This also applies to the puller, which in far-field can also point towards the wall, as it is well-known.
Very close to the wall, where lubrication applies, a weak puller is upright and a strong one tilted against the wall normal. 
Finally, the pusher under lubrication points upward or, if it is sufficiently strong, also towards the wall, where it is then pinned to 
the wall. In the far field it tends towards the well-known parallel orientation ($\vartheta^{*} \rightarrow \pi/2$ for $z\rightarrow \infty$).
Thus, when leaving the wall, the pusher has to tilt away from the normal and then slides along the wall. This gives a first understanding of the sliding state, illustrated in Fig.\ \ref{fig:hist}. Of course, one also has to show the existence of a stable sliding height, which we will do in Sec.\ \ref{floating_heights}.
The recurrent floating state of the pusher has an upward orientation, which is not stable. Thus, it can only be a transient state.

In Ref.\ \cite{LintuvuoriMarenduzzo2016} the authors also provide a matched expansion, where they extrapolate between the lubrication and the far-field regime. In particular, this approach describes how the stable orientation of a pusher tilts from $\vartheta^{*} = 0$ towards $\pi/2$ when swimming away from the wall.

\section{Discussion of squirmer states}
\label{sec.discussion}
\subsection{Simulation results}

\begin{figure*}
\begin{center} 
	\resizebox{\textwidth}{!}{
	\includegraphics{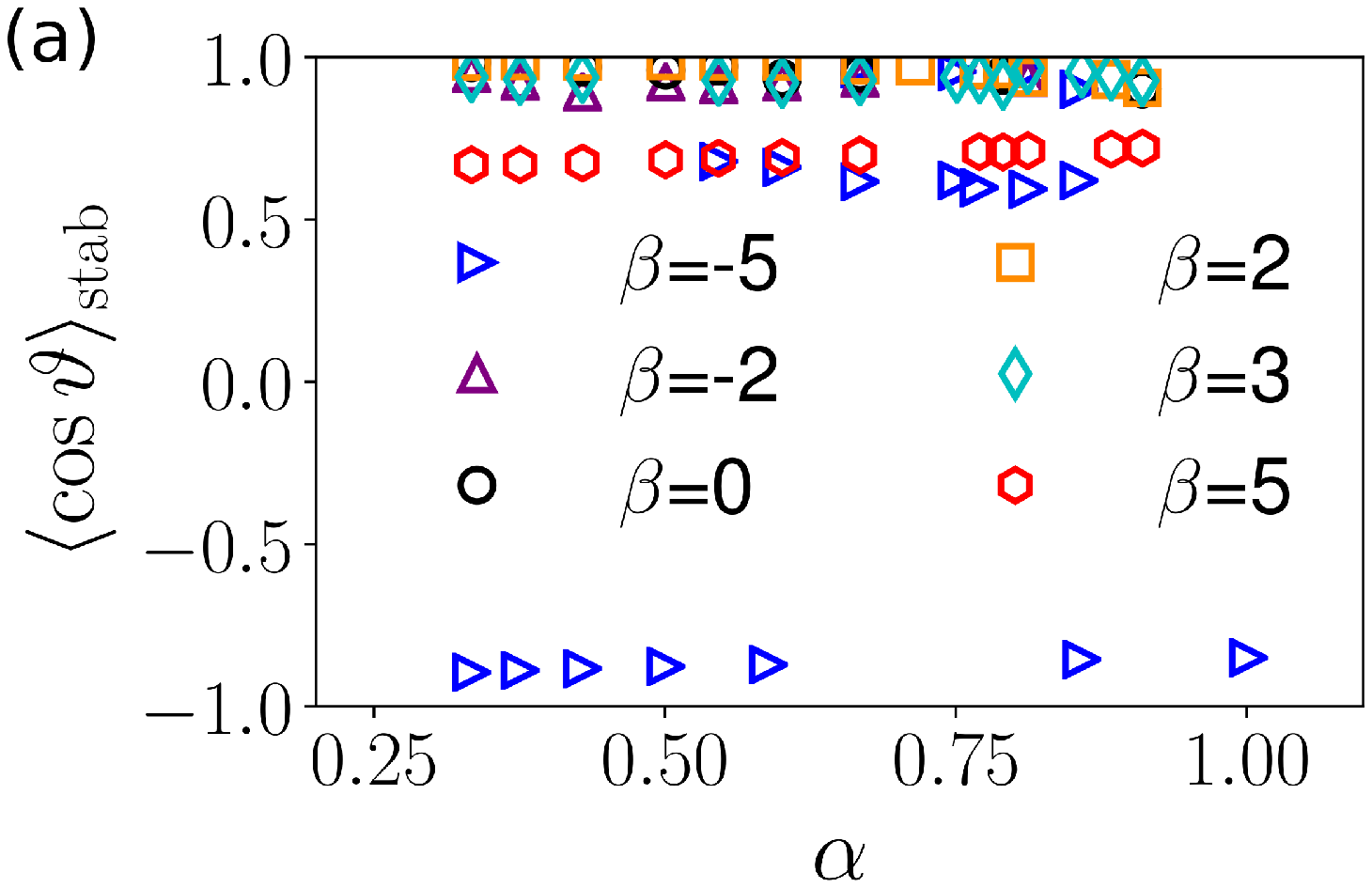}
	\includegraphics{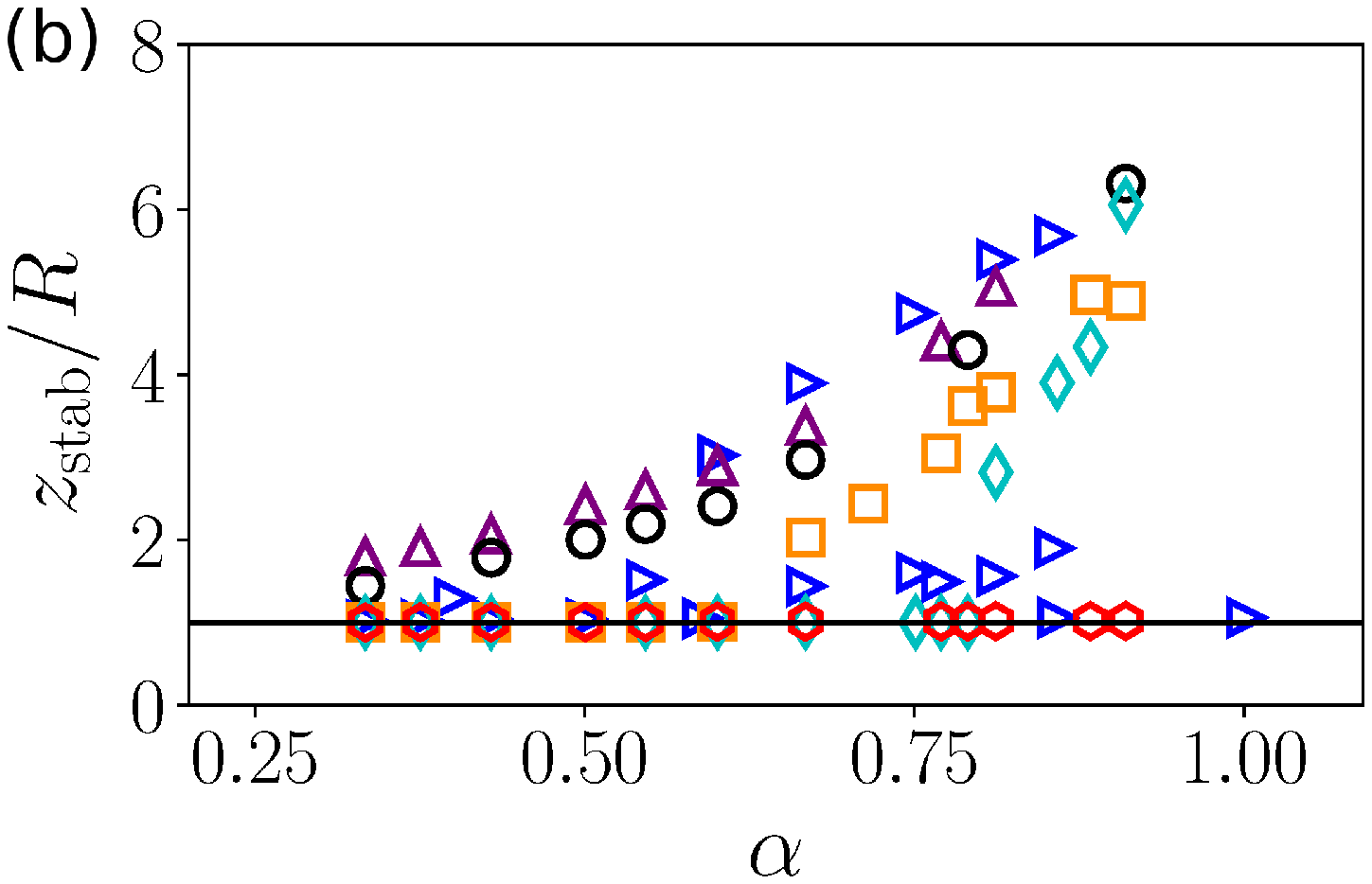}
    \includegraphics{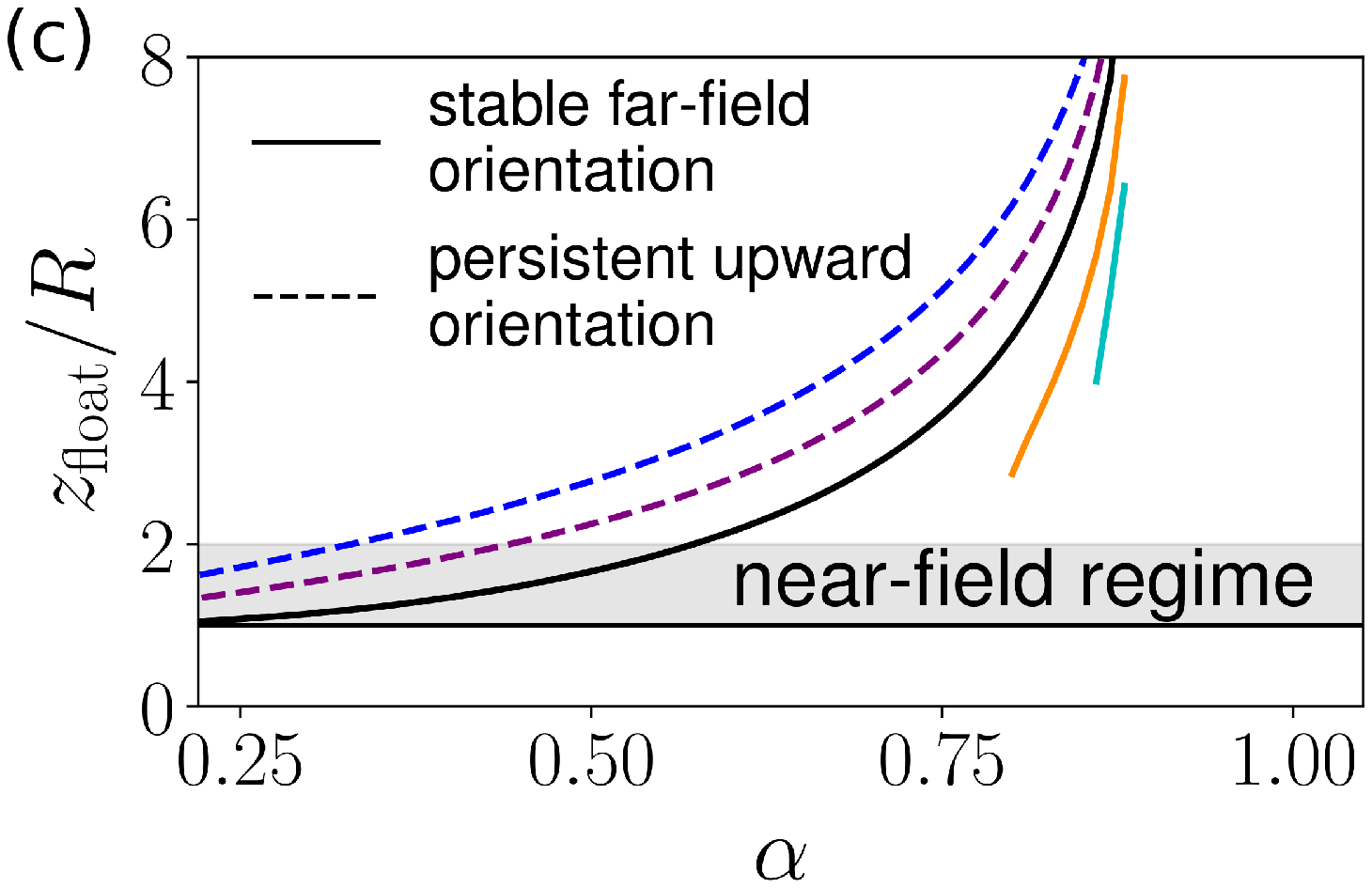}}
\end{center}
   \caption{From simulations: mean stable orientation $\langle \cos \vartheta \rangle_{\mathrm{stab}}$ (a) and stable height $z_{\mathrm{stab}}$ (b) plotted versus $\alpha = v_0 /v_g$ for different $\beta$. Depending on the observed motional state, we plot in (b) floating, sliding, and wall-pinned heights. 
   (c) From theory: floating height $z_{\mathrm{float}}$ versus $\alpha$, determined in far-field approximation for $\vartheta^{\ast} = 0$. Solid line: $\vartheta^{\ast} = 0$ is stable upward orientation; dashed line: $\vartheta^{\ast} = 0$ is an unstable, equilibrium orientation or fixed point ($\Omega_2 =0$).  Note the colors in (a) - (c) refer to the same squirmer type as indicated in (a).}
   \label{fig:stable_heights}
\end{figure*}

We simulated single squirmers under gravity varying both the squirmer parameter $\beta$ and the velocity ratio $\alpha = v_0 / v_g$ of the swimming and the bulk sedimentation velocity. In Sec.\ \ref{subsec.phenom} we already explained that cruising trajectories between the bottom and top wall occur for $\alpha > 1$ due to
the persistent motion at high P\'eclet numbers. If $\alpha \ll 1$, gravity dominates and the squirmer simply sinks  to the bottom wall. For intermediate values, $0.2 < \alpha < 1$,  and depending on $\beta$, we find constant floating, recurrent floating and wall sliding, as well as wall-pinned states, which we already introduced shortly in Sec.\ \ref{subsec.phenom}. In the following we describe these states in more detail. In Figs.\ \ref{fig:stable_heights}(a) and (b) we show an overview of our numerical results by plotting the mean stable orientation $\langle \cos \vartheta \rangle_{\mathrm{stab}}$ and the observed 
(multi)stable heights $z_{\mathrm{stab}}$ versus $\alpha$ for different $\beta$.
\footnote{While we plot the maximum value of the floating height, the corresponding heights of the sliding and wall-pinned state are shown as an average over 
time restricted to the respective state.}
These quantities are characteristic for the different squirmer states.

\begin{figure}
\centering
\resizebox{0.45\textwidth}{!}{  
\includegraphics{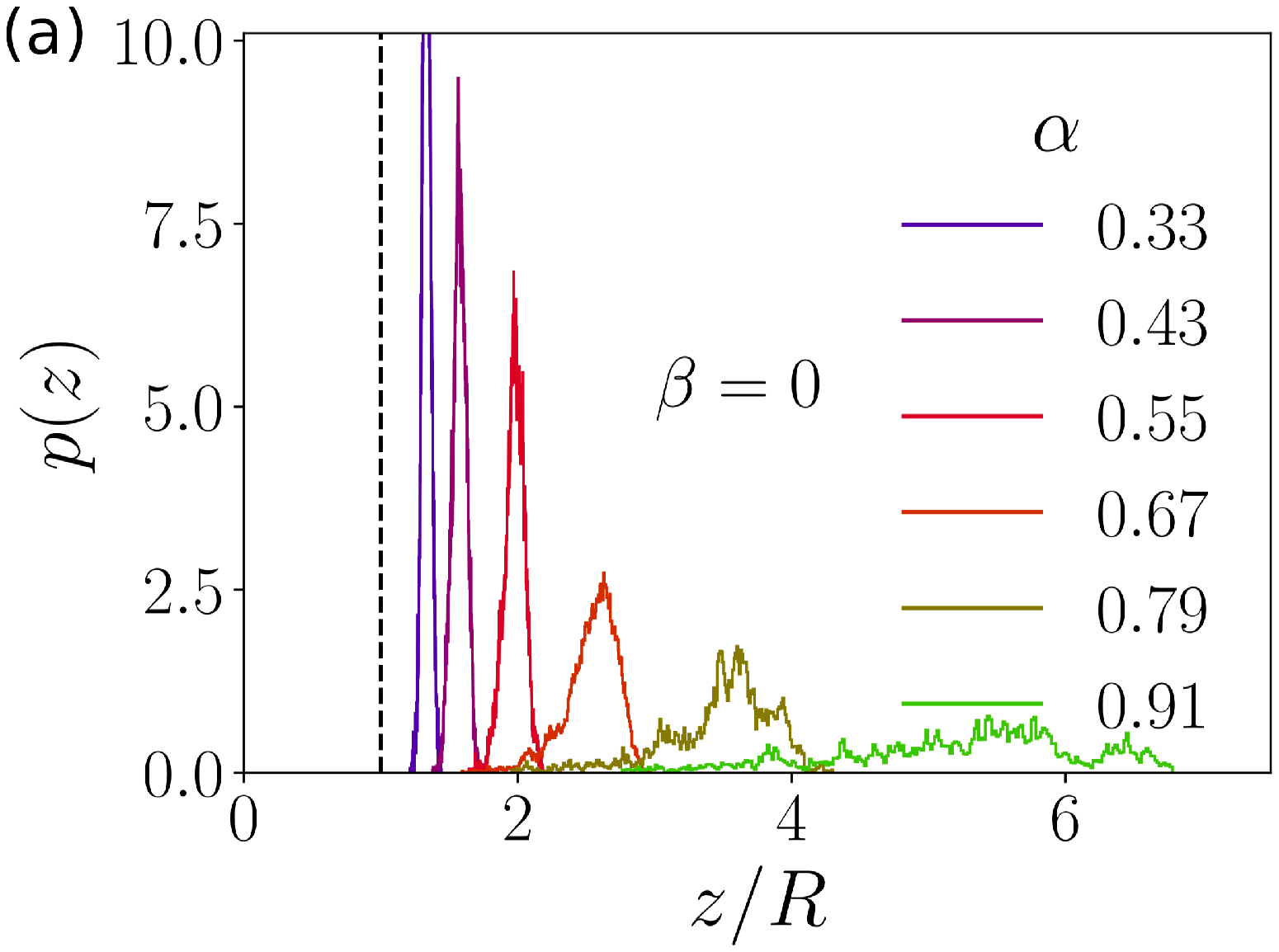}}
\resizebox{0.45\textwidth}{!}{ 
\includegraphics{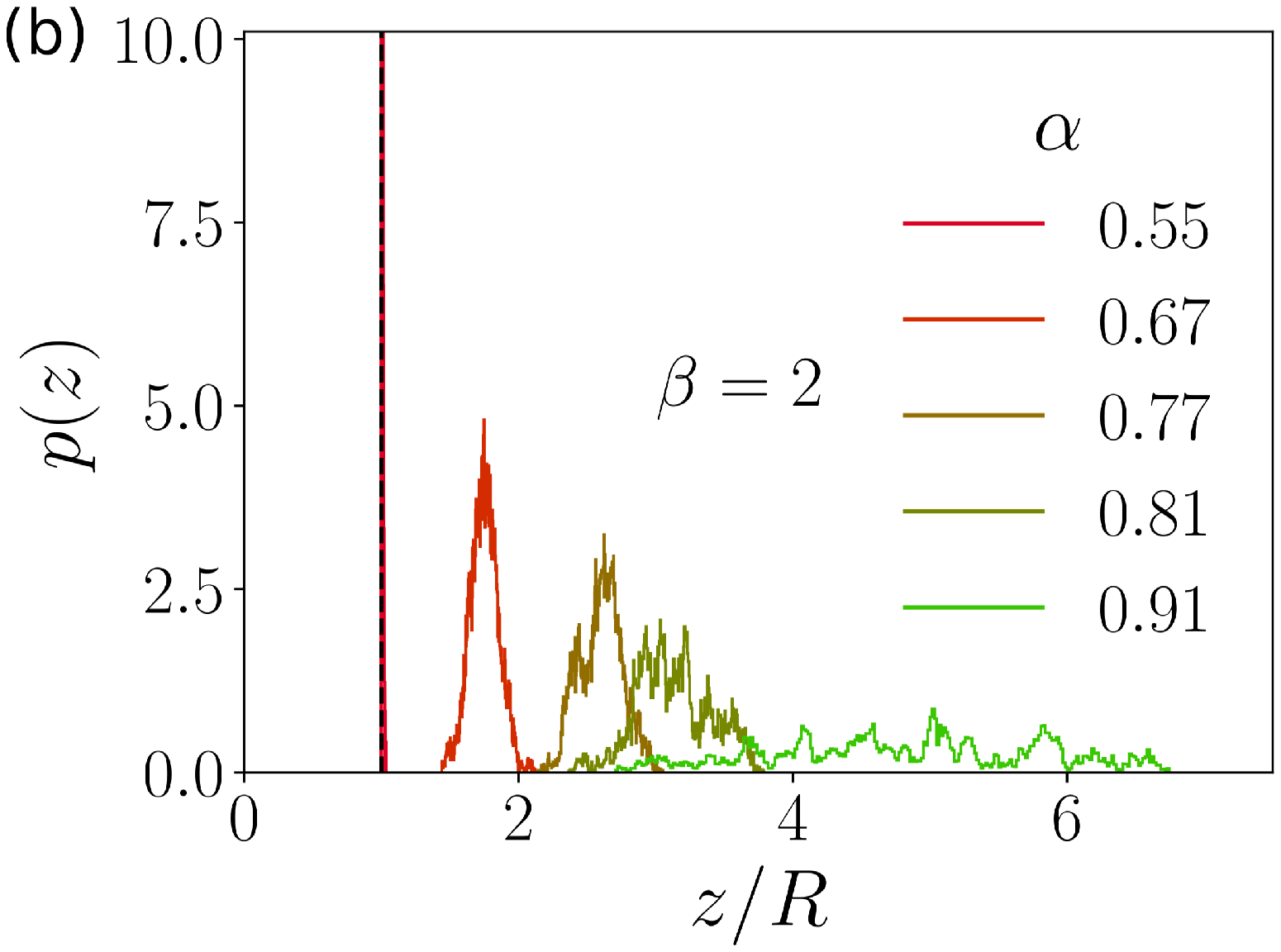}}
\caption{
Distribution of squirmer heights $p(z)$ for different $\alpha$ 
for 
(a) floating neutral squirmers and 
(b) pullers with $\beta = 2$.
}
\label{histbeta02}
\end{figure}

\subsubsection*{1. Constant floating above the wall}

Neutral squirmers float at a finite height above the wall for intermediate $\alpha$. This is nicely illustrated in Fig.\ \ref{histbeta02}(a), where we plot the height distribution for different $\alpha$.
The floating height continuously shifts away from the bottom wall with increasing $\alpha$. 
We plot its maximum value in Fig.\ \ref{fig:stable_heights}(b).

In addition, the height fluctuations increase with $\alpha$ indicated by the 
growing width of the height distributions.
As explained in Sec.\ \ref{subsec.stable_orient}, the neutral squirmer assumes an upward orientation, which is also visible in Fig.\ \ref{fig:stable_heights}(a). However, thermal fluctuations tilt the squirmer and, as a result, it sinks down. 
This generates the height distributions. They become broader with increasing $\alpha$, since at larger floating heights the restoring torque on the squirmer orientation is smaller. Nevertheless, the orientational stabilization means that after a downward excursion the swimmer regains its floating height rather quickly.

Figure\ Fig.\ \ref{histbeta02}(b) shows that pullers also float,
however, only if $\alpha$ exceeds a certain threshold value $\alpha_\mathrm{th}$.
The maximum floating height plotted in Fig.\ \ref{fig:stable_heights}(b) for $\beta = 2$ and $\beta = 3$ illustrates the threshold value, which increases with the squirmer parameter $\beta$. As a consequence, we  do not observe any floating 
for the strong puller with $\beta=5$. It is pinned to the wall with a tilted orientation [see Fig.\ \ref{fig:stable_heights}(a), $\langle \cos \vartheta \rangle_{\mathrm{stab}} < 1$]. Thus, the threshold value $\alpha_\mathrm{th}$ 
separates wall-pinned states from floating states.

\subsubsection*{2. Recurrent floating and sliding}

The pusher's behavior is rather different. From the height distribution in
Fig.\ \ref{fig:hist} (a) we clearly see that its dynamical state is bistable. Sometimes it resides at the wall and sometimes above the wall.
It floats recurrently. During floating phases the pusher floats at systematically larger heights than the neutral squirmer, as demonstrated in Fig.\ \ref{fig:stable_heights}(b), in particular, for $\beta = -5$. 
However, while the height of neutral squirmers and pullers during floating is recovered after a disturbance in the upward orientation, strong pushers sink down 
towards the wall and assume their sliding state. 

We already know from Sec.\ \ref{subsec.stable_orient} that the upward orientation of a pusher during floating is not stable, while we argued that the tilted orientation at smaller heights should be stable [see also $\langle \cos \vartheta \rangle_{\mathrm{stab}}  < 1$ for $\beta=-5$ in Fig.\ \ref{fig:stable_heights}(a)]. 
Occasionally, fluctuations in the orientation vector towards $\cos \vartheta = 0$ [see Fig.\ \ref{fig:stable_heights}(a)] let the squirmer 
rise to its floating height since reorientation either by thermal fluctuations or angular drift proceeds slowly. As can be seen in Fig.\ \ref{fig:stable_heights}(b), strong pushers ($\beta=-5$) do not show recurrent floating for $\alpha \lesssim 0.6$ and pushers with $\beta =-2$ do not assume the sliding state. We discuss this 
further in Sec.\ \ref{floating_heights}.

\subsubsection*{3. Wall-pinned states}

\begin{figure}
\centering
\resizebox{0.45\textwidth}{!}{ \includegraphics{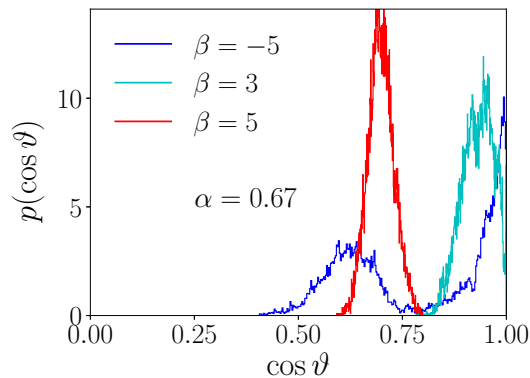}}
\caption{Distribution $p(\cos\theta)$ of orientation $\cos \theta$ during motion of a squirmer for $\alpha = 0.67$ 
and different squirmer types $\beta$ for recurrent floating and sliding ($\beta=-5$) and in the wall pinned state ($\beta=3,5$).
}
\label{fig:hist_orientation}
\end{figure}

Both pushers and pullers also assume a state, where they are pinned to the wall and do not manage to leave it during the whole simulation time. For the puller this state occurs for $\alpha < \alpha_\mathrm{th}$ and the orientation is roughly vertical with $0.6  < \langle\cos\vartheta\rangle_{\mathrm{stab}} < 1$ depending on $\beta$ [see Fig.\ \ref{fig:stable_heights}(a)].
Note that the observed angles of pullers in the simulations (see also Fig.\ \ref{fig:hist_orientation}) do not quantitatively recover 
the stable orientations of lubrication theory in Tab. \ref{tab:orientations}, which would give $\langle \cos \vartheta \rangle = \frac{1}{3} $ 
and $\frac{1}{5}$ for $\beta = 3$ and $5$, respectively.
Possible reasons for the deviation are that the squirmer does not always sit exactly at the wall due to thermal flucutations and that
we cannot expect MPCD to quantitatively resolve the lubrication result at the wall.

The pushers, however, occupy a separate state, where they point towards the wall [see $\langle\cos\vartheta\rangle_{\mathrm{stab}} \approx -1$ in Fig.\ \ref{fig:stable_heights}(a)], which is in agreement with the stable near-field orientation in Tab.\ \ref{tab:orientations}. 
It is not impossible that a transition between the recurrent floating state and the wall-pinned state occurs eventually, although we never observed it within the simulation time.

\subsection{Stable floating and sliding heights}
\label{floating_heights}

\begin{figure}
\centering
\resizebox{0.40\textwidth}{!}{ \includegraphics{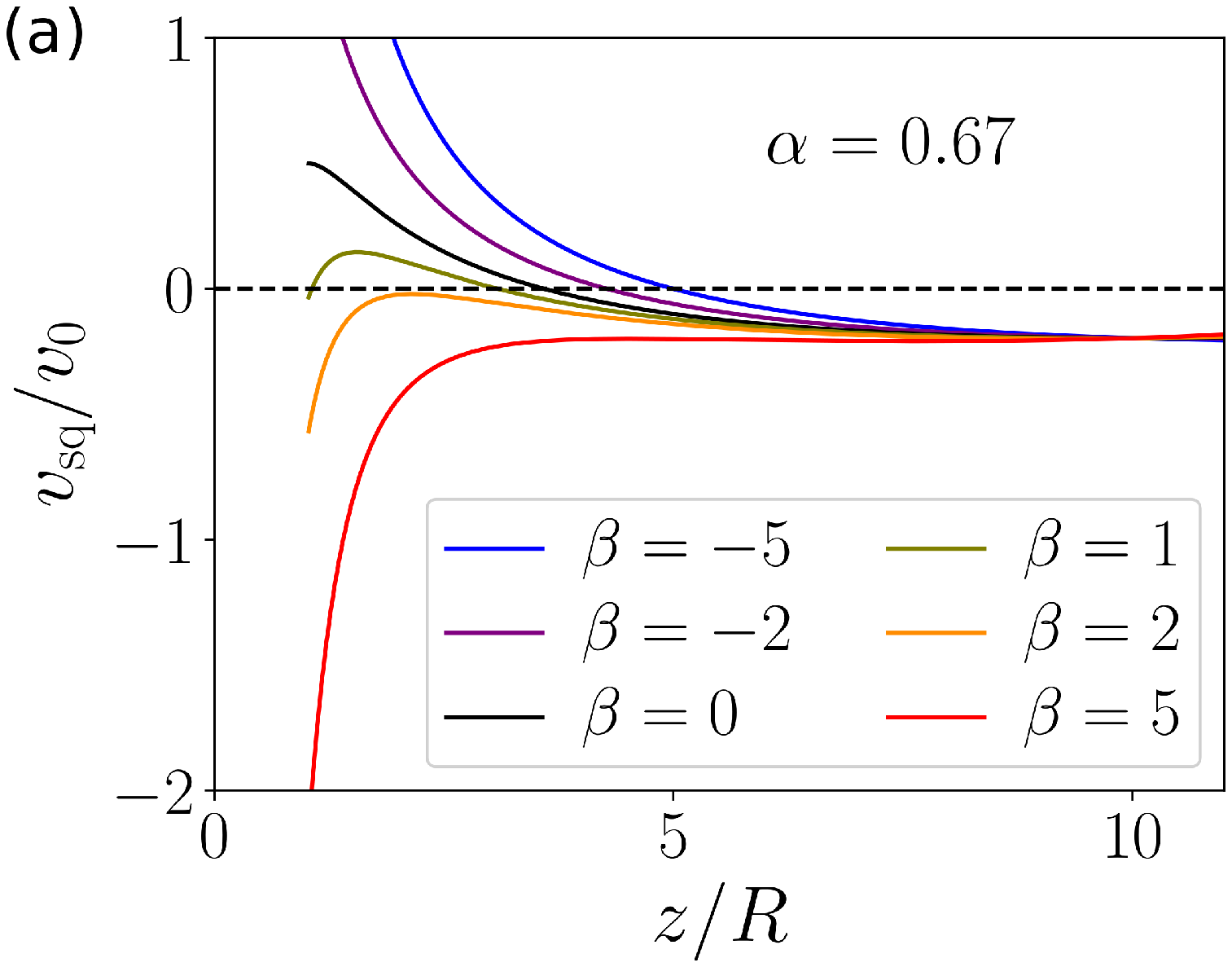}}
\resizebox{0.40\textwidth}{!}{ \includegraphics{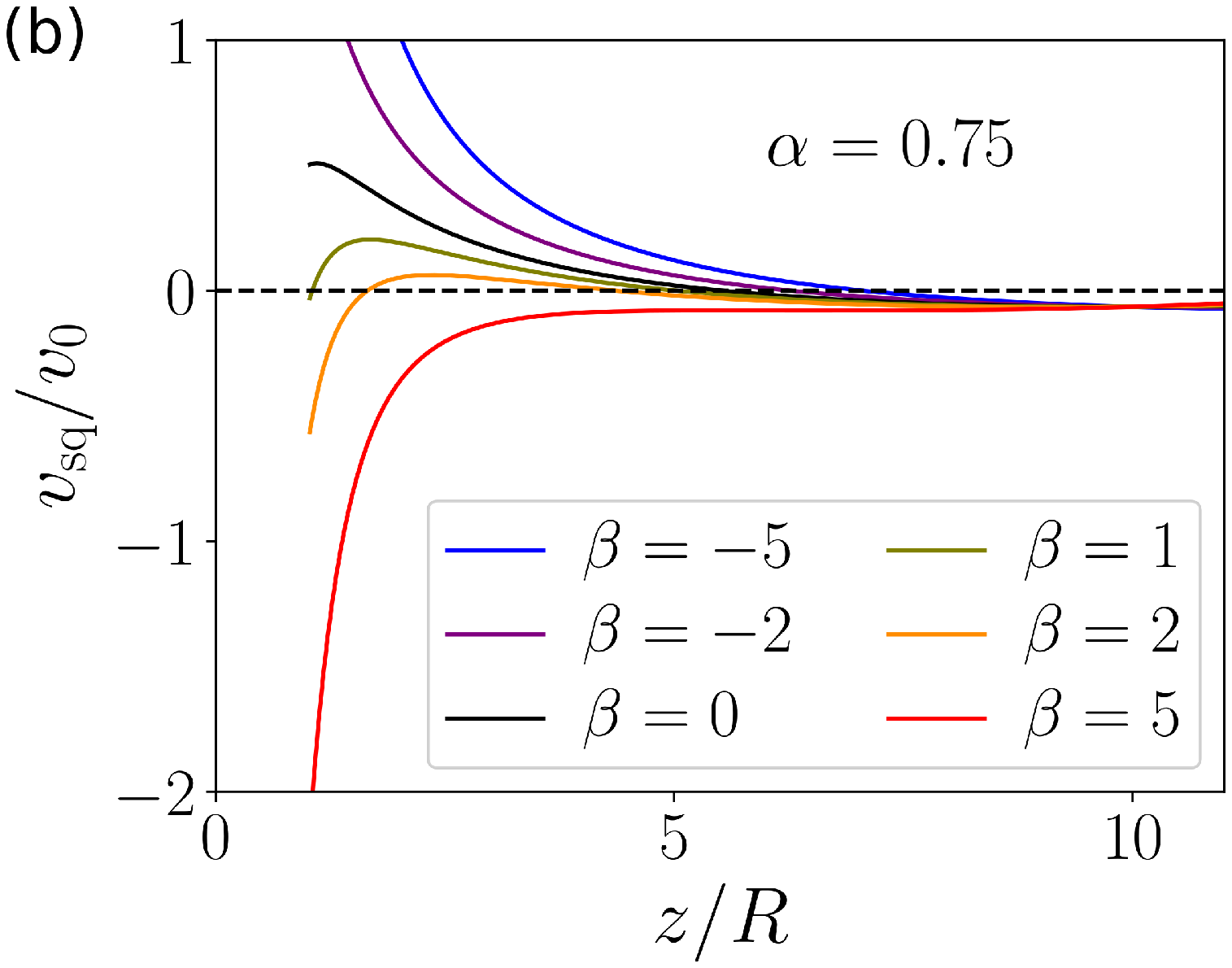}}
\caption{Vertical squirmer velocity $v_{\mathrm{sq}}$ versus height $z$ for $\alpha=0.67$ (a) and $\alpha=0.75 $ (b). The far-field approximation of eq.\ (\ref{eq:velocity_ff}) 
is used for vertical orientation $\theta^* = 0$ and different squirmer parameters $\beta$.}
\label{fig:vel_stab_ff}
\end{figure}

It remains to analyze the vertical squirmer velocity of eq.\ (\ref{eq:velocity_ff}) in order to determine the floating heights.
Neutral squirmers and pullers float with upward stable orientation. Thus we set $\vartheta^* = 0$ in eq. (\ref{eq:velocity_ff}) and
plot $v_{\mathrm{sq}}$ versus $z$ for different squirmer types $\beta$ in Fig.\ \ref{fig:vel_stab_ff} for $\alpha = 0.67$ (a) and $\alpha = 0.75$ (b).
A stable floating height $z_\mathrm{float}$ is determined by $v_{\mathrm{sq}} = 0$ and $d v_{\mathrm{sq}} / dz < 0$. Such heights 
always exist for the neutral squirmer. The corresponding curve in Fig.\ \ref{fig:stable_heights}(c) shows that $z_\mathrm{float}$ continuously 
increases with $\alpha$ as observed in the simulations [see Fig.\ \ref{fig:stable_heights}(b)].

For the pullers the behavior is different. This can be nicely illustrated for $\beta=2$ in Fig.\ \ref{fig:vel_stab_ff}. For $\alpha=0.67$ (a),
the squirmer velocity is always negative and the puller sinks down to the wall. However, increasing $\alpha$ to 0.75 (b), a stable
floating height develops, which explains the existence of a threshold value $\alpha_{\mathrm{th}}$ above which the puller starts to float.
The resulting floating heights for $\beta = 2$ and 3 are drawn in Fig.\ \ref{fig:stable_heights}(c). One realizes that $\alpha_{\mathrm{th}}$ increases with $\beta$ as observed in the simulations.

\begin{figure}
\centering
\resizebox{0.45\textwidth}{!}{ \includegraphics{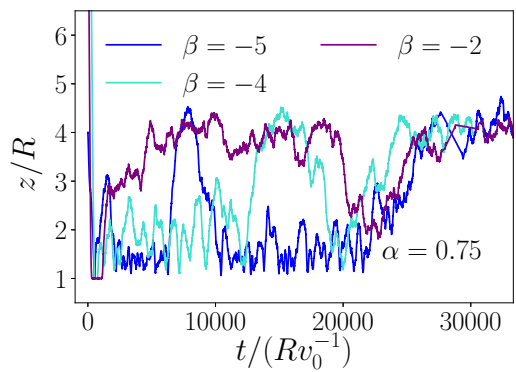} }
\caption{Height variations $z(t)$ for pushers with $\beta=-5$, $-4$, and 
$-2$ at $\alpha =0.75$. Larger floating and smaller sliding heights are distinguishable.
}
\label{fig:traj_bimodal}
\end{figure}

We already stated that the pusher does not have a stable upward orientation besides when it is at the wall, where it always swims upwards.
As already discussed, the pusher assumes the sliding state with a stable tilted orientation, which keeps it from swimming too high.
Instead, due to strong orientational fluctuations  (see left peak in the orientational distribution function for $\beta=-5$
in Fig.\ \ref{fig:hist_orientation}), it performs a strong irregular 
up-and-down movement close to the wall (see video M3 and Fig.\ \ref{fig:traj_bimodal}).
Weak pushers reach larger sliding heights compared to strong pushers (see sliding heights in Fig.\ \ref{fig:traj_bimodal})
since their sliding angles tend towards the stable upward orientation of the neutral squirmer and is thus smaller.

Finally, when orientational fluctuations in the sliding state drive the pusher towards an upward orientation, it will move upwards as the positive vertical velocity $v_{\mathrm{sq}}$ for small $z$ shows in Fig.\ \ref{fig:vel_stab_ff}.
Ultimately, it reaches its floating height at $v_{\mathrm{sq}} = 0$.
Due to the large directional persistence of the squirmer in our simulations, it keeps floating for a considerable amount of time until orientational fluctuations strongly tilt the pusher's orientation. As a results, it sinks down, enters the sliding state, and the cycle begins again. In Fig.\ \ref{fig:stable_heights}(c) we plot the floating heights of the recurrent floating state. They nicely compare to the simulation results in  Fig.\ \ref{fig:stable_heights}(b). In particular, the recurrent floating height is larger for stronger pushers. Note that 
the pusher's recurrent floating state corresponds to a saddle point in the dynamical system of eq.\ (\ref{eq:dynamical_system}) since the upright orientation is only an unstable fixed point.

In Fig.\ \ref{fig:stable_heights}(b) we observe that at small $\alpha$ the strong pusher ($\beta = -5$) does not assume the 
recurrent floating state. Due to the stronger gravity, the squirmer is closer to the wall and thus reorientation towards the vertical is hindered by a larger restoring torque in the sliding state. 
For weak pushers ($\beta = -2$ in Fig.\ \ref{fig:traj_bimodal}) the difference in recurrent floating and sliding heights becomes smaller and tends to zero for the neutral squirmer. Thus we did not attempt to determine and plot sliding heights in Fig.\ \ref{fig:stable_heights}(b).

\section{Conclusion}
\label{sec.conclusion}
A single squirmer under gravity is conceptually simple, yet in our study we could classify very variable microswimmer dynamics at high P\'{e}clet numbers.
The decisive factors for the observed motional states are hydrodynamic interactions with the no-slip surface, gravity, and thermal noise, which are usually present in experimental systems. 
Since in experiments one can vary density mismatch between fluid and a non-neutrally buoyant particle, as well as temperature, particle radius, and also active velocity, we expect a wide range of values for the ratio $\alpha$ of the swimming and bulk-sedimentation velocity to be experimentally accessible.
Our study thus provides an interesting example for the non-equilibrium dynamics of a microswimmer, in particular in the regime where sedimentation velocity and active velocity become similar. 
 
\begin{figure}
\centering
\resizebox{0.45\textwidth}{!}{
 \includegraphics{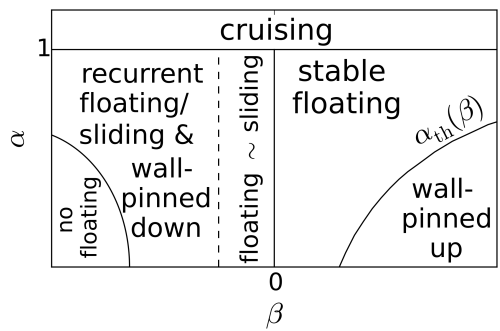}
}
\caption{Schematic representation of the motional states of a squirmer depending on the position in parameter space $\alpha$ versus $\beta$.
}
\label{fig:diagram_states}
\end{figure}

At $\alpha > 1$ we observe a cruising state, where the neutral squirmer 
and puller swim between the upper and lower bounding wall due to 
their large persistence while pushers stay at the walls.
In contrast, at $\alpha < 1$ several motional states occur depending on squirmer type $\beta$ and reduced swimming speed $\alpha$. While neutral squirmers constantly float above the wall with upright orientation, pullers float for $\alpha$ larger than a threshold value $\alpha_{\mathrm{th}}$ and are pinned to the wall below $\alpha_{\mathrm{th}}$. The threshold value increases with $\beta$. In contrast, pushers show recurrent floating with upright orientation due to their strong orientational persistence, while they also slide along the wall at lower heights, which is the stable state. For weak pushers it is difficult to distinguish between both states since for $\beta \rightarrow 0$ they both tend towards the floating state of the neutral squirmer. At small $\alpha$ strong pushers do not show recurrent floating due to the strong wall-induced restoring torques, which keeps them in the stable sliding state. Finally, pushers are also able to exhibit a wall-pinned state with downward orientation. 
We summarize our findings about the motional states in a schematic diagram $\alpha$ versus $\beta$ in Fig.\ \ref{fig:diagram_states}.

To arrive at the full understanding of the phenomenology of our MPCD simulations, we performed a theoretical anaysis of the total vertical squirmer velocity and its rotational velocity. Both are strongly determined by wall-induced linear and angular velocities due to the hydrodynamic interactions of the squirmer flow fields with the wall and thus depend on the squirmer type $\beta$.
The floating and sliding states correspond to stable fixed points in the height and orientation of the squirmer, while the upward orientation in the recurrent floating state is only transient and occurs due to the strong persistent swimming.
 
We plan to advance this research by including an external torque acting on the swimmers, e.g., due to their bottom-heaviness. 
Such a system has been studied in~\cite{WolffStark2013} without any hydrodynamics.
Interestingly, for large swimming speeds and strong bottom-heaviness
inverted sedimentation profiles occur.
We will also drastically increase the particle number, similar to Ref.\ \cite{KuhrStark2017}, where we expect these inverted profiles to become unstable due to hydrodynamic interactions between the squirmers.

Another interesting research direction are  catalytically powered microswimmers \cite{GolestanianAjdari2007,MoranPosner2017,YangRipoll2014}.
Their phoretic fields also interact with bounding walls. This changes the surface flow fields on the microswimmers and thereby their translational and rotational velocities. This setup has already attracted much attention \cite{SimmchenSanchez2016,UspalTasinkevych2015,UspalTasinkevych2015rheotaxis,Crowdy2013,IbrahimLiverpool2015,MozaffariMaldarelli2016}.

\ack
We would like to thank Andreas Z\"ottl, Corinna C. Maass and William Uspal for stimulating discussions
and the referees for their stimulating comments.
This project was funded by Deutsche Forschungsgemeinschaft  through the priority program SPP 1726 (grant number STA352/11) and the research training group GRK 1558.

\section*{References}

\bibliography{lit_NJP}

\bibliographystyle{iopart-num}

\end{document}